\documentstyle[twoside,aps,prl,multicol,epsf]{revtex} 
\begin{document}
\def\B.#1{{\bbox{#1}}}
\def\BC.#1{{\bbox{\cal{#1}}}}
\title{{\rm PHYSICAL REVIEW LETTERS \hfill
Submitted }\\
Dissipative Scaling Functions in Navier-Stokes Turbulence: 
Experimental Tests} 
\author {Adrienne L. Fairhall, Victor S. L'vov and
Itamar Procaccia}
\address{Department
of~~Chemical Physics, The Weizmann Institute of Science,
Rehovot 76100, Israel}
\maketitle
\begin{abstract}
  A recent theoretical development in the understanding of the
  small-scale structure of Navier-Stokes turbulence has been the
  proposition that the scales $\eta_n(R)$ that separate inertial from
  viscous behavior of many-point correlation functions depend on the
  order $n$ and on the typical separations $R$ of points in the
  correlation. This is a proposal of fundamental significance in
  itself but it also has implications for the inertial range scaling
  behaviour of various correlation functions.  This dependence has
  never been observed directly in laboratory experiments.  In order to
  observe it, turbulence data which both display a well-developed
  scaling range with clean scaling behaviour and are well-resolved in
  the small scales to well within the viscous range is required.  We
  have analysed the data of the experiments performed in the
  laboratory of P. Tabeling of Navier-Stokes turbulence in a helium
  cell with counter-rotating disks, and find that this data satisfies
  these criteria. We have been able to find strong evidence for the
  existence of the predicted scaling of the viscous scale.
\end{abstract}

\begin{multicols}{2}
\newpage

The familiar approach to the statistical theory of Navier-Stokes
turbulence \cite{Fri} concentrates on the properties of two-point 
differences  of the Eulerian velocity field ${\bf u}({\bf x},t)$ and
their moments, termed structure functions:
\begin{equation}
S_n(R) = \langle | \B.u(\B.r + \B.R) - \B.u(\B.r) |^n \rangle\ .
\end{equation}
In isotropic homogeneous turbulence, these structure functions are
observed to behave as a power-law in $R$, $S_n(R) \sim R^{\zeta_n}$,
with scaling exponents $\zeta_n$ that may be universal.  This scaling
holds within a range of scales between the outer scale $L$ determined
by the system size or the forcing, and some inner scale $\eta$
determined by the viscosity, below which the velocity field is
essentially smooth. In this regime $S_n(R) \sim R^n$.  The usual
definition of the viscous scale was established by Kolmogorov from the
balance of the viscosity $\nu$ and the mean energy flux
$\bar{\epsilon}$, according to $\eta \sim L {\rm
  Re}^{-1/(2-\zeta_2)}$.  The observation of intermittency in
turbulence is contrary to the relatively straightforward picture which
follows from Kolmogorov's assumption of a single length scale, and
suggests that a more complex situation applies. We begin by a review
of the derivation of the functional behaviour of the viscous scale
(found in more detail in \cite{LP-3}).

One possible approach to the question of cross-over scales follows a
recent theoretical trend to concentrate on the more general
simultaneous many-point correlation functions ${\cal F}_n$ of velocity
differences rather than on two-point quantities only.  These are
defined in terms of two-point differences $ {\B.w}({\B.r},{\B.r}',t)
\equiv \B.u (\B.r',t)- {{\bf u}}(\B.r,t), $ as
\begin{eqnarray} &&\B.{\cal
F}_n({\B.r}_1,{\B.r}'_1;{\B.r}_2,{\B.r}'_2; \dots;{\B.r}_n,{\B.r}'_n)
\nonumber \\
&=& \langle \B.w({\B.r}_1,{\B.r}'_1)
\B.w({\B.r}_2,{\B.r}'_2) \dots
\B.w({\B.r}_n,{\B.r}'_n) \rangle \,,
\label{defF}
\end{eqnarray}
where $\langle \cdot \rangle$ denotes averaging, and all coordinates
are distinct. Time labels have been dropped as only simultaneous
correlations will be considered.  Homogeneous scaling here means that
\begin{equation}
{\bf {\cal F}}_{n}(\lambda\B.r_1,
\lambda\B.r'_1,\!\dots\! ,\lambda\B.r'_{n})
=\lambda^{\zeta_{n}}\!\B.{\cal F}_{n}
(\B.r_1,\B.r'_1, \!\dots\!,\B.r'_{n})\,,
\label{hom}
\end{equation}
with $\zeta_n$ the scaling exponent.  Taking the time derivative of
$\B.{\cal F}_{n}$, using the Navier-Stokes equations to evaluate each
$\partial \B.u /\partial t$ and considering the stationary state
where $\partial{\cal F} /\partial t = 0$, one derives the following
statistical balance equation:
\begin{equation}
{\cal D}_n(\B.r_1,\B.r_1';...;\B.r_n,\B.r_n') = 
\nu {\cal J}_n(\B.r_1,\B.r_1';...;\B.r_n,\B.r_n')\  .
\end{equation}
Here the term ${\cal D}_n$ arises from the nonlinear interaction
term and may be written as
\begin{eqnarray}
\label{dn}
 && {\cal D}_n^{\alpha_1 \alpha_2 ... \alpha_n} 
(\B.r_1,\B.r_1';...;\B.r_n,\B.r_n') 
= 
\int d \B.r \sum_{j=1}^n P_{\alpha_j \beta} (\B.r)
\nonumber \\ &\times &
\langle 
w_{\alpha_1}(\B.r_1,\B.r_1') \dots
L^\beta(\B.r_j,\B.r_j',\B.r) \dots
w_{\alpha_n}(\B.r_n,\B.r_n')
\rangle\,,  \\ \nonumber
&&
L^\beta(\B.r_j,\B.r_j',\B.r)  
\equiv 
\frac{1}{n}
\sum_{k = 1}^{n} \Big[ w_{\gamma}(\B.r_j - \B.r,
\B.r_k)\nabla_j^\gamma
\nonumber \\
&+& w_{\gamma}(\B.r'_j - \B.r,\B.r'_k)
\nabla_{j'}^\gamma \Big] 
w_\beta(\B.r_j - \B.r,\B.r'_j - \B.r)\ . \nonumber
\end{eqnarray}
In the above $P_{\alpha_j \beta} (\B.r)$ is the projection operator.
The RHS with coefficient $\nu$, the kinematic viscosity,
results from the viscous term and is defined
\begin{eqnarray}
& & {\cal J}_n(\B.r_1,\B.r_1';...;\B.r_n,\B.r_n')  =
\sum_{j=1}^n (\nabla^2_j + \nabla^2_{j'} ) \nonumber \\
& \times & \langle w_{\alpha_1}(\B.r_1,\B.r_1') \dots
w_{\alpha_j}(\B.r_j,\B.r_j') \dots
w_{\alpha_n}(\B.r_n,\B.r_n')
\rangle\ .
\end{eqnarray}
This equation provides the means to determine the scale of the viscous
range. The balance equation expresses the competition between the
small-scale viscous effects and the interesting non-linear dynamics,
and intuitively, the viscous scale should be the scale at which the
two effects become comparable. This raises a rather subtle question.
As there is a balance of the two terms, it appears that the scaling
properties of the correlators must be determined by the viscous term.
However one believes that the properties of the inertial range
quantities are {\em independent} of the details of the viscous range.
This apparent paradox can be understood if one considers the
separations of the coordinates of the correlation functions. If all
separations are in the inertial range, there are no small-scale
quantities, and the contribution of ${\cal J}_n$ will be negligible,
leaving a homogeneous equation ${\cal D}_n = 0$. Non-trivial scaling
can arise from special solutions for the terms in the sum in ${\cal
  D}_n$ that exactly cancel one another.  Now as some coordinates in
the correlation functions approach one another, the gradients in
${\cal J}_n$ will begin to show their effect: these pick up the
smallest separation $r_{\rm min}$, introducing a factor of $1/r_{\rm
  min}^2$. As $r_{\rm min} \rightarrow 0$, this term is no longer
negligible; the scaling solution of the homogeneous inertial range
equation will no longer be valid and one obtains the smooth viscous
result.

Thus one wishes to estimate the terms of the balance equation both
in the inertial range and in the limit where some coordinates 
approach one another, in order to observe this crossover and 
estimate its scalelength. 

It can be shown \cite{LP-3} that in the case where all separations 
are of order $R$ one may evaluate ${\cal D}$ simply as 
\begin{equation}
{\cal D}_n \sim S_{n+1}(R)/R\ .
\end{equation}
This can be demonstrated by proving that the integral in (\ref{dn})
converges in both limits, so that the typical evaluation at $R$ is 
correct. As points in the correlation approach one another, this
evaluation remains valid; but cancellations between terms will
no longer occur.

The second term ${\cal J}_n$ can be estimated directly as
$
{\cal J}_n(R) \sim S_n(R)/R^2\ .
$
In the limit when some separation $r_{ij} \rightarrow 0$, this
evaluation is replaced by
$
{\cal J}_n(r_{ij};R) \sim {\cal F}_n(r_{ij}; R)/r_{ij}^2\,,
$
where ${\cal F}_n(r_{ij}; R)$ is shorthand notation for ${\cal F}_n$ 
with an overall typical separation $R$ and some pair of coalescing
points of smaller separation $r_{ij}$.
Now the balance equation gives 
\begin{equation}
{\cal F}_n(r_{ij}; R) \sim r_{ij}^2 S_{n+1}(R)/\nu R\ .
\end{equation}
This gives the evaluation of ${\cal F}_n(r_{ij}; R)$ 
for a small separation in the viscous regime.

Now we wish to compare this with an evaluation for 
${\cal F}_n(r_{ij};R)$ when the small distance is still in the
inertial range.  To do so we invoke the fusion rules derived in 
\cite{LP-1,LP-2}.
These rules predict the behaviour of multipoint correlation functions 
as some pairs of coordinates approach one another, or ``fuse''. 
The essential result concerns a correlation of $n$ pairs of points
${\BC.F}_n$ where $p$ pairs of coordinates
$\B.r_1,\B.r'_1\dots \B.r_p,\B.r'_p$, $(p<n)$ of $p$ velocity
differences coalesce, with typical separations between the  
coordinates
$|\B.r_i -\B.r'_i| \sim r$ for $i \le p$, and where all 
other separations  are of the order of $R$, $r\ll R \ll L$.  
Let us denote such a
correlation as ${\BC.F}_n^{(p)}$.
In a homogeneous isotropic
scaling system, the fusion rules predict
\begin{eqnarray}\label{fusion1}
& & {\BC.F}_n^{(p)}
({\B.r}_1,{\B.r}'_1;
\dots;{\B.r}_n,{\B.r}'_n)
\\ &= &
{\tilde{\BC.F}}_p({\B.r}_1,{\B.r}'_1;
\dots;{\B.r}_p,{\B.r}'_p)
{\bf\Psi}_{n,p}
({\B.r}_{p+1},{\B.r}'_{p+1};\dots;{\B.r}_n,{\B.r}'_n) \ , 
\nonumber
\end{eqnarray}
where ${\tilde{\BC.F}}_p$ is a tensor of rank $p$ associated with the
first $p$ tensor indices of ${\BC.F}_n$, and it has a homogeneity
exponent $\zeta_p$.  The $(n-p)$-rank tensor ${\bf
  \Psi}_{n,p}({\B.r}_{p+1},{\B.r}'_{p+1};\dots;{\B.r}_n,\B.x'_n)$ is a
homogeneous function with a scaling exponent $\zeta_n-\zeta_p$, and is
associated with the remaining $n-p$ indices of ${\BC.F}_n$. In terms
of the scaling of structure functions, this can be expressed for $p$
points coalescing to a distance $r$ and all other points with typical
separation $R$ as (abbreviating the coordinate dependence of
${\BC.F}_n^{(p)}$)
\begin{equation}
{\BC.F}_n^{(p)}(r;R)
\sim S_p(r) S_n(R)/S_p(R)\ .
\end{equation}
In the special case that $p = 1$, due to the vanishing of the
average of a single difference in isotropic turbulence, the
leading order result is
\begin{equation}
{\BC.F}_n^{(1)}(r;R)
\sim S_2(r) S_n(R)/S_2(R).
\end{equation}
Applying this result to the correlation we previously denoted
${\cal F}_n(r_{ij};R)$ one obtains
\begin{equation}
{\cal F}_n(r_{ij};R) \sim S_2(r_{ij}) S_n(R)/S_2(R).
\end{equation}

Thus we have an inertial range and a viscous range evaluation of
${\cal F}_n(r_{ij};R)$. Let us take the scale $\eta_n$ 
to be that at which the two evaluations coincide. Balancing in
the two-point case one recovers the Kolmogorov estimate, 
$\eta_2 \sim L {\rm Re}^{-1/(2-\zeta_2)}$.
For other values of $n$ one finds 
\begin{equation}
\label{eta}
\eta_n(R) = \eta_2\left( \frac{R}{L} \right)^{x_n}, \ 
x_n = \frac{\zeta_n + \zeta_3 - \zeta_{n+1} - \zeta_2}{2 - \zeta_2}\ .
\end{equation}

In order to test this proposition experimentally, we will consider
direct measurements of the function ${\cal J}_n(R)$.  To make a
comparison with the one-dimensional data obtained from experiments we
take a form defined by
\begin{equation}
J_n(\rho;R)= \left<\tilde \nabla^2_\rho {\B.u}(\B.r) 
\left[{\B.w}(\B.r,\B.r+{\B.R})\right]^{n-1} \right> 
\cdot{\B.R}/R .
\label{Jn}
\end{equation}
For discrete data, the Laplacian operator $\tilde \nabla^2_\rho$ 
in (\ref{Jn}) is taken to be a second order finite difference 
of longitudinal components of the velocity,
\begin{equation}
\tilde \nabla^2_\rho {\bf u}(\B.r) = [{\B.w}(\B.r,\B.r
+\B.\rho)- {\B.w}(\B.r,\B.r-\B.\rho)]
\cdot\B.\rho/\rho^3. \label{surr} \end{equation}
From the discussion above, one expects a different scaling 
for $\rho$ above and below the dissipative scale. For $\rho$ 
in the inertial range, the estimation of (\ref{fusion1}) is applicable 
and one predicts
\begin{equation}
J_n(\rho;R) = C_nJ_2(\rho) S_n(R)/2S_2(R) \ , \quad \rho\gg\eta\ . 
\label{bigrho}
\end{equation}
where $C_n$ is an $R$-independent dimensionless
constant which may have $n$-dependence.
However, for $\rho$ in the viscous regime, 
\begin{equation}
J_n(\rho;R) = \tilde C_n J_2(\rho) S_{n+1}(R)/S_3(R) \ , \quad 
\rho\ll\eta \ , \label{smallrho}
\end{equation}

\narrowtext
\begin{figure}
\epsfxsize=9.0truecm
\epsfbox{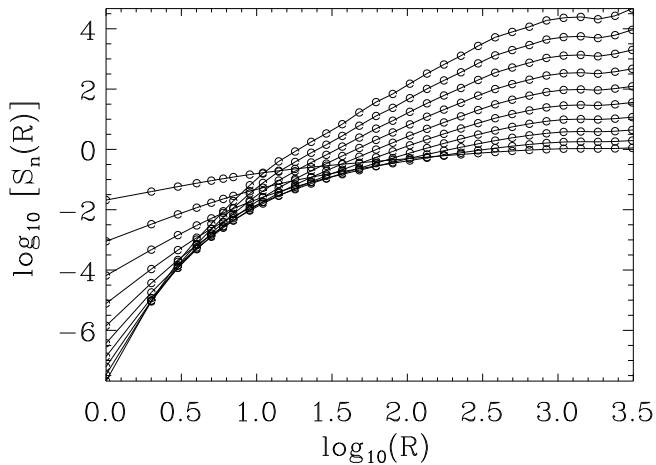}
\caption
{Log-log plot of the structure functions $S_n(R)$ as a function of  
$R$
for $n=1-10$.}
\label{Fig1}
\end{figure}
\begin{figure}
\epsfxsize=9.0truecm
\epsfbox{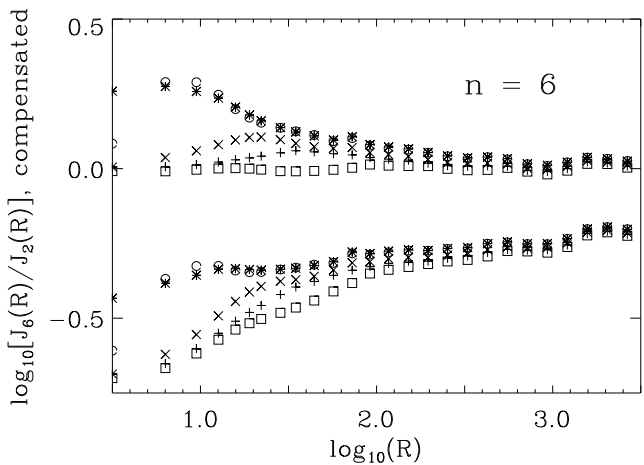}
\caption
{Log-log plot of the normalised function $J_4(R)/J_2(R)$ as a function 
of $R$ for $\rho = $1,3,7,11 and 29, represented by $\Box$,
$+$, $\times$, $*$ and $\circ$ respectively. The data are
compensated in the upper plots by the inertial range fusion rule
prediction $S_4(R)/S_2(R)$ and in the lower by the result in the
viscous regime $S_5(R)/S_3(R)$.}
\label{j4}
\end{figure}
\noindent
where $\tilde C_n$ is some other coefficient.
One can show that $J_2$ is equal to the mean dissipation 
$\langle |\nabla u(x)|^2 \rangle$, and is thus expected to be 
$R$-independent.
The explicit prefactor containing $\rho$ is included in $J_2$;
we will consider only the $R$ dependence resulting from the scaling
of (\ref{eta}).

These predictions are tested in data obtained by F. Belin and H.
Williame in the laboratory of P. Tabeling at Ecole Normale Superieure;
see for example \cite{96BTW,94ZTMW}.  The data are time signals of the
velocity field taken from a low-temperature cell of helium gas
enclosed in a cylinder and driven by counter-rotating disks.  The
helium is maintained at a constant temperature around 5K and at a
controlled pressure. The recordings were made using a hot-wire probe
consisting of a 7$\mu$m carbon fibre coated with evaporated gold apart
from an active area of size of order 10$\mu$m.  The frequency response
of the probe can range between 10 and 50 kHz.  The data had very long
acquisition times, containing up to 30 million samples.  The resulting
statistics are well-resolved and stationary.  The Taylor microscale
Reynolds number and the Kolmogorov microscale $\eta$ were determined
through the usual procedure of surrogating time for space (by Taylor's
hypothesis), and data is available for a range both of $R_\lambda$ and
of minimum resolved lengthscale $r/\eta$.  We selected data sets
according to the small-scale resolution, and although the $R_\lambda$
was not extremely high, a distinct inertial range is evident.  The
data presented here has a minimum resolved distance $r/\eta$ of 1.18
and $R_\lambda$ of 418.

In Fig.~\ref{Fig1} we present the structure functions $S_n(R)$ as a
function of $R$.  In all figures, spatial separations have units of
sampling times, and the velocity is normalised by the RMS velocity.
This figure shows that we have one and a half decades of ``inertial
range'' (between, say 10 and $500$ sampling units) and that the
lengthscales below 10 units are smooth and well-resolved.  The initial
logarithmic slope of the $n$th structure function at 1 unit is close
to $n$ (deviating successively more for higher order $n$, as would be
expected). There is reasonably well-defined scaling behaviour up to
order 10.

Our aim is to try to expose the postulated cross-over in scaling
behaviour in $R$ of $J_n(\rho;R)$ as a function of $\rho$. We have
calculated the correlation functions $J_n(\rho;R)$ for several values
of $\rho$ from the minimum distance of 1 unit up to a value well into
the inertial range.  Note that the difference in scaling that is
expected is rather small; one expects the scaling exponent of
$J_n(\rho;R)$ to cross over from $\zeta_n - \zeta_2$ to
$\zeta_{n+1}-\zeta_3$, which for the usual values of scaling exponents
obtained in turbulent experiments gives a difference of the order of
0.15 for $n = 4$ to 0.2 for $n = 8$.  Thus we do not present the
results in terms of calculated exponents, as one cannot justifiably
separate values of this closeness on the basis of exponents calculated
on a limited inertial range. Instead we will examine the function as a
whole.

In Figs.2-4 we display the results. The three figures show
$J_n(\rho;R)$ for a single value of $n$, for $n = $ 4, 6 and 8.  For
each $n$ there are results for five values of $\rho$, $\rho = $1, 3,
7, 11 and 29. The figures each show two sets of data, one in which the
calculated $J_n$s have been compensated by the inertial range
prediction $J_2(\rho) S_n(R)/S_2(R)$ (the upper set of functions), and
the second showing the same data compensated by the dissipative range
result $J_2(\rho) S_{n+1}(R)/S_3(R)$.  Hence in the upper set we
expect to see that for inertial range values of $\rho$, the resulting
plots are constant in $R$ in the inertial range.  In principle as
there is no knowledge of the coefficient $C_n$, the value of the
constant $C_n$ can be different for different $n$.  (It is trivially 1
for $n = 2$.) We hope to see that the dissipative range scaling is a
better fit as $\rho \rightarrow 0$.

\begin{figure}
\epsfxsize=9.0truecm
\epsfbox{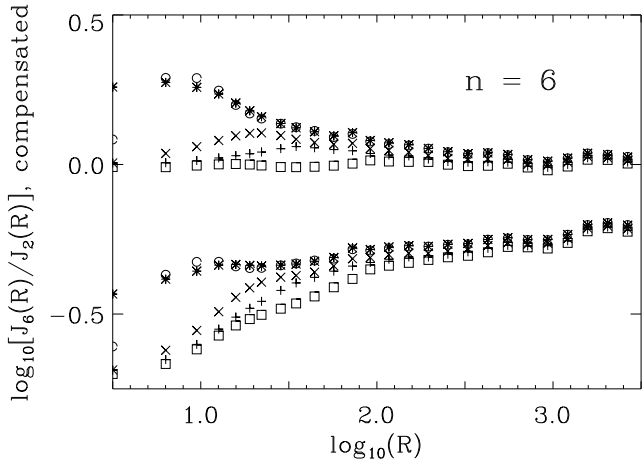}
\caption
{Log-log plot of the normalised function $J_6(R)/J_2(R)$ as a function 
of $R$ for $\rho = $1,3,7,11 and 29, represented by $\Box$,
$+$, $\times$, $*$ and $\circ$ respectively,
compensated in the upper plots by $S_6(R)/S_2(R)$ and in the lower by 
$S_7(R)/S_3(R)$.}
\label{j6}
\end{figure}
\narrowtext
\begin{figure}
\epsfxsize=9.0truecm
\epsfbox{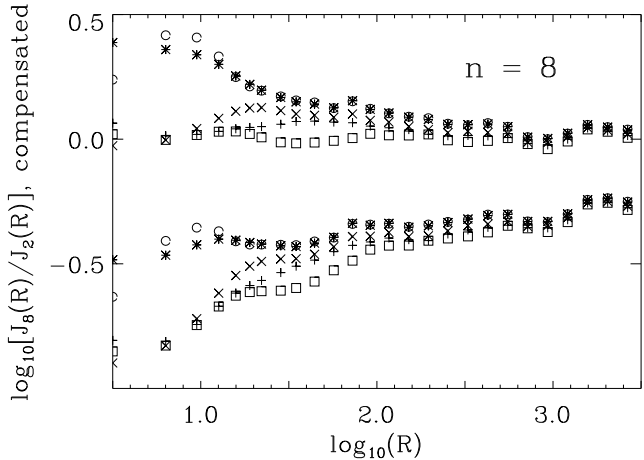}
\caption
{Log-log plot of the normalised function $J_8(R)/J_2(R)$ as a function 
of $R$ for $\rho = $1,3,7,11 and 29, represented by $\Box$, $+$, 
$\times$, $*$ and $\circ$ respectively,
compensated in the upper plots by $S_8(R)/S_2(R)$ and in the lower by 
$S_9(R)/S_3(R)$.}
\label{j8}
\end{figure}
The upper sets of plots in each figure show that for the inertial
range values of $\rho$, the inertial scaling (\ref{bigrho}) is very
well realised.  This scaling has been previously observed in
turbulence data \cite{sreeni} but the agreement in this data is more
impressive: it has smaller fluctuations, and the agreement continues
into the viscous scales, which has not previously been seen to be the
case. Comparing between the figures, for different values of $n$ one
finds that all coefficients $C_n$ are very near to 1.  Comparing
different values of $\rho$, there is a continuous dependence on $\rho$
in the functional behaviour of $J_n(\rho;R)$.  There is a clear
deviation from the inertial range scaling as $\rho$ decreases, and the
smallest value of $\rho$ shows a small but definite slope. The plots
corrected by the dissipative range scaling show a distinct indication
that there is a tendency toward this slope. The effect becomes more
apparent for larger $n$ as the separations between the two scalings
becomes larger. These functions of course also show larger statistical
fluctuations.

One should note that the plots for $\rho = 1$ and $\rho = 3$ are
almost identical. This shows that in fact the data is only resolved
well to $\rho = 3$, and no further information is gained in the
subsequent refinements of scale. Thus the minimum lengthscale
that can be considered as resolved is in fact of order $3\eta$. 
It is therefore not surprising that a clean scaling of (\ref{smallrho})
is not precisely observed. Nonetheless the trend in 
that direction is clearly visible.

We have been able to find convincing experimental evidence that 
the viscous scale of $n$-point multipoint correlation functions is an 
anomalous scaling function. We have verified the inertial range
fusion rules and given evidence that the small scale structure behaves
according to the theoretical predictions. 

\acknowledgments 
This work has been supported in part
by the TAO Exchange Program (ALF), the German Israeli 
Foundation, the US-Israel Bi-National Science Foundation and the 
Naftali and Anna Backenroth-Bronicki Fund for Research in Chaos and
Complexity. We thank Patrick Tabeling, 
Frederic Belin and Herve Williame for providing us with their data, and ALF
thanks them and their coworkers at Ecole Normale Superieure 
for their hospitality.

\end{multicols}

\begin{references}

\bibitem{Fri} U. Frisch.
{\em Turbulence: The Legacy of A.N. Kolmogorov} 
(Cambridge University Press, Cambridge, 1995). 

\bibitem{LP-3}
V.S. L'vov and I. Procaccia, Phys. Rev. Lett. {\bf 77}, 3541 (1996). 

\bibitem{LP-1}
V.S. L'vov and I. Procaccia, Phys. Rev. Lett. {\bf 76}, 2896 (1996). 

\bibitem{LP-2}
V.S. L'vov and I. Procaccia, Phys. Rev. E, {\bf 54}, 6268 (1996).

\bibitem{96BTW}
F. Belin, P. Tabeling and H. Williame
Physica D {\bf 93}, 52 (1996) 

\bibitem{94ZTMW}
G. Zocchi, P. Tabeling, J. Maurer and H. Williame
Phys. Rev. E {\bf 50}, 5, 3693 (1994) 

\bibitem{sreeni} 
A. Fairhall, B. Dhruva, V. S. L'vov, I. Procaccia and K. R. Sreenivasan,
Phys. Rev. Lett., in press.

\bibitem{SY}
V. Yakhot and Y. Sinai, Phys. Rev. Lett., {\bf 63}, 1963 (1991) 

\bibitem{96CLPP}
E.S.C. Ching, V.S. L'vov, E. Podivilov and I. Procaccia, 
Phys. Rev. E, {\bf 54}, 6364 (1996).

\bibitem{97FGLP} 
A. Fairhall, B. Galanti, V. S. L'vov, and I. Procaccia,
{\em Direct numerical simulations of the Kraichnan model}, 
Phys. Rev. Lett., submitted.

\end{references}
\end{document}